\documentclass[%
 aip,
 amsmath,amssymb,
 reprint,%
]{revtex4-1}

\usepackage{graphicx}
\usepackage{dcolumn}
\usepackage{bm}

\usepackage[utf8]{inputenc}
\usepackage[T1]{fontenc}
\usepackage{mathptmx}
\usepackage{etoolbox}

\makeatletter
\def\@email#1#2{%
 \endgroup
 \patchcmd{\titleblock@produce}
  {\frontmatter@RRAPformat}
  {\frontmatter@RRAPformat{\produce@RRAP{*#1\href{mailto:#2}{#2}}}\frontmatter@RRAPformat}
  {}{}
}%
\makeatother
\begin{document}

\preprint{AIP/123-QED}

\title[Keldysh crossover in one-dimensional Mott insulators]{Keldysh crossover in one-dimensional Mott insulators}
\author{Kazuya Shinjo}
\affiliation{
Computational Quantum Matter Research Team, RIKEN Center for Emergent Matter Science (CEMS), Wako, Saitama 351-0198, Japan
}
\author{Takami Tohyama}
\affiliation{%
Department of Applied Physics, Tokyo University of Science, Katsushika, Tokyo 125-8585, Japan
}%

\date{\today}

\begin{abstract}
Recent advancements in pulse laser technology have facilitated the exploration of non-equilibrium spectroscopy of electronic states in the presence of strong electric fields across a broad range of photon energies. 
The Keldysh crossover serves as an indicator that distinguishes between excitations resulting from photon absorption triggered by near-infrared multicycle pulses and those arising from quantum tunneling induced by terahertz pulses.
Using time-dependent density-matrix renormalization group, we investigate the emergence of the Keldysh crossover in a one-dimensional (1D) Mott insulator.
We find that the Drude weight is proportional to photo-doped doublon density when a pump pulse induces photon absorption.
In contrast, the Drude weight is suppressed when a terahertz pulse introduces doublons and holons via quantum tunneling.
The suppressed Drude weight accompanies glassy dynamics with suppressed diffusion, which is a consequence of strong correlations and exhibits finite polarization decaying slowly after pulse irradiation.
In the quantum tunneling region, entanglement entropy slowly grows logarithmically.
These contrasting behaviors between the photon-absorption and quantum tunneling regions are a manifestation of the Keldysh crossover in 1D Mott insulators and provide a novel methodology for designing the localization and symmetry of electronic states called subcycle-pulse engineering.
\end{abstract}

\maketitle

%

\section{\label{sec:level1}Introduction}

The photoexcited nonequilibrium dynamics of Mott insulators has been actively studied both theoretically and experimentally, providing valuable insights into nontrivial quantum many-body phenomena in strongly correlated electron systems. 
Recent advancements in pulse lasers have enabled the investigation of non-equilibrium spectroscopy of electronic states under strong electric fields across a wide range of frequencies.

At low photon energy $\Omega$, doublons and holons are induced through nonlinear dielectric processes triggered by DC electric fields and high-field terahertz pulses.
On the other hand, at high photon energy $\Omega$, they are induced through optical processes triggered by a multicycle pulse in a near-infrared band. 
In other words, a low-$\Omega$ pulse leads to quantum tunneling~\cite{Oka2003, Oka2005, Oka2008, Oka2010, Eckstein2010, Oka2012}, while a high-$\Omega$ pulse leads to multi-photon absorption~\cite{Takahashi2008, Filippis2012, Lenarcic2014, Eckstein2016, Hashimoto2016, Shinjo2017, Miyamoto2019}.
The Keldysh crossover~\cite{Keldysh1965} distinguishes these two excitation processes.

In this paper, we show that optical properties in one-dimensional (1D) extended Hubbard model (1DEHM) differ significantly between excitations by high-$\Omega$ pulses and low-$\Omega$ pulses, which induce photon absorption and quantum tunneling, respectively.
Using density-matrix renormalization group (DMRG), we find that a transient absorption spectrum induced by photon absorption shows the enhancement of the Drude weight proportional to photo-doped doublon density, while that by quantum tunneling shows no enhancement of the Drude weight.
The localized nature of excited states induced by a low-$\Omega$ pulse is associated with finite polarization, which persists after pulse irradiation and slowly decays.
The induced polarization breaks inversion symmetry, which is monitored by the second-harmonic generation.

Additionally, when doublons and holons are induced by quantum tunneling, the entanglement entropy slowly grows logarithmically, which indicates the localized nature of excited states.
Our findings suggest that photon absorption in 1D Mott insulators leads to a Mott insulator-to-metal transition, while quantum tunneling results in a Mott insulator-to-glass transition.

\section{\label{sec:level2}Model: One-dimensional extended Hubbard model}

\begin{figure}[t]
\includegraphics[width=0.45\textwidth]{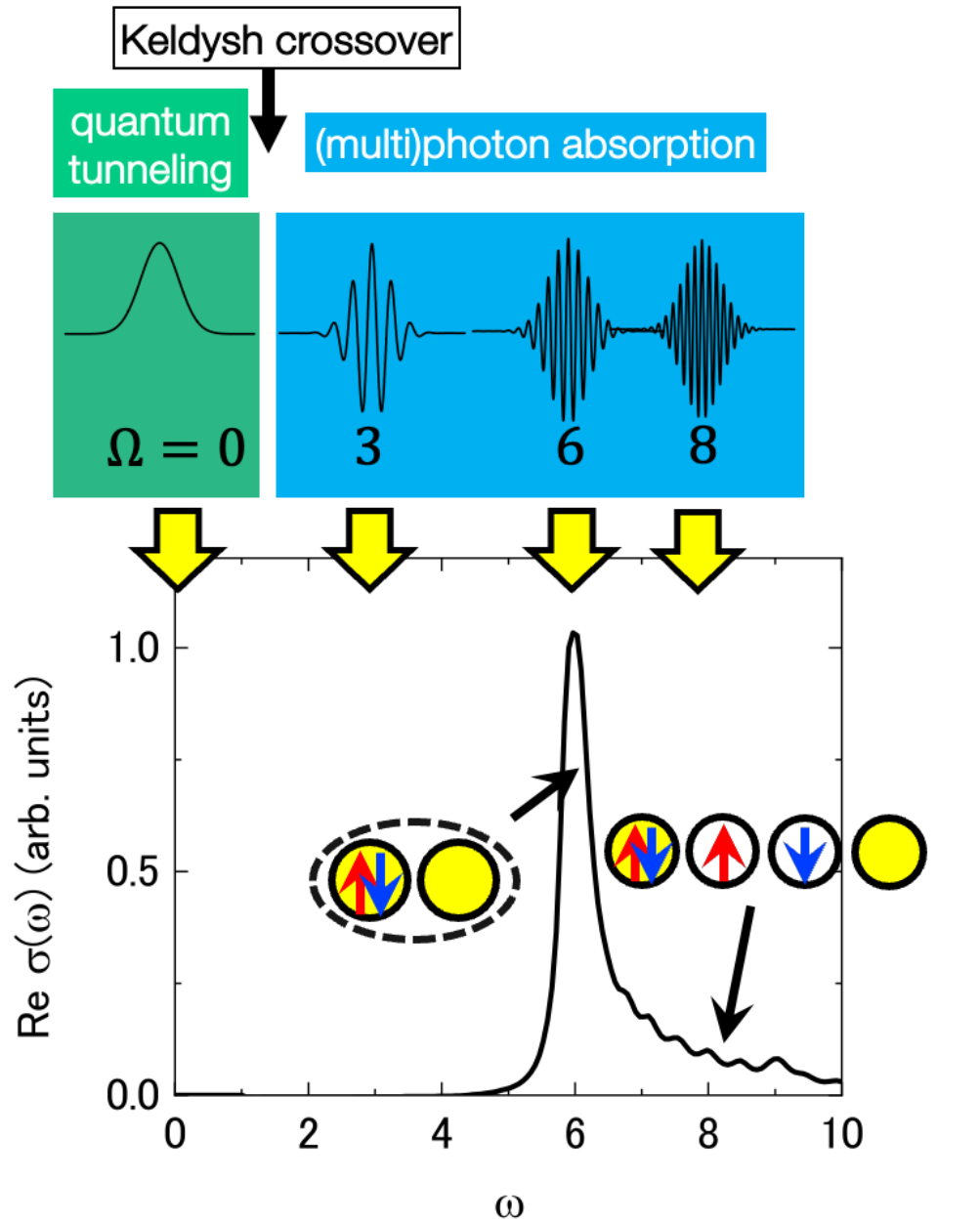}
\caption{A schematic picture of the scope of this paper.
Linear absorption spectrum Re$[\sigma(\omega)]$ of an $L=32$ half-filled 1DEHM with $(U,V)=(10,3)$ is shown.
There is an excitonic level at $\omega=6$, where a doubly occupied site called doublon and an empty site called holon are bounded, followed by a continuum with separated doublon and holon.
Pulse shapes with $\Omega=0$, 3, 6, and 8 are also shown.
The yellow arrows indicate the corresponding photon energies for each pulse.
}
\label{Fig1}
\end{figure}

Electronic states of 1D Mott insulators such as ET-F$_2$TCNQ and Sr$_2$CuO$_3$ are well-described by a half-filled 1DEHM with on-site Coulomb interaction $U$ and nearest-neighbor repulsive Coulomb interaction $V$, whose Hamiltonian reads
\begin{align}
    \mathcal{H}_\text{EH}=-t_\text{h} \sum_{i, \sigma}\left( c_{i,\sigma}^{\dag}c_{i+1,\sigma}+\text{H.c.} \right)
+U\sum_{i}n_{i,\uparrow}n_{i,\downarrow} 
+ V\sum_{i} n_{i}n_{i+1},
\end{align}
where $c_{i,\sigma}^{\dag}$ is the creation operator of an electron with spin $\sigma (= \uparrow, \downarrow)$ at site $i$, and $n_i=\sum_\sigma n_{i,\sigma}$ with $n_{i,\sigma}=c^\dagger_{i,\sigma}c_{i,\sigma}$.
We take hopping energy $t_\mathrm{h}$ to be the unit of energy ($t_\mathrm{h}=1$) in the following discussion.
Experimentally, $t_\text{h} \simeq 0.1$eV and $t_\text{h} \simeq 0.3 $eV for ET-F$_2$TCNQ~\cite{Wall2011} and Sr$_2$CuO$_3$~\cite{Kim2006, Schlappa2012}, respectively.

In 1D Mott insulators, the kinetic motion of holons and doublons is not fundamentally influenced by spin degrees of freedom. 
This phenomenon is commonly known as spin-charge separation. 
The distinct features of charge dynamics in 1D Mott insulators become evident in an absorption spectrum upon doping.
Specifically, when carriers are introduced in 1D Mott insulators, it is typical for spectral weights above the Mott gap to dynamically redistribute towards those at $\omega=0$, i.e., the Drude weight.

The optical conductivity, responsible for a linear absorption spectrum, has been obtained using the dynamical DMRG~\cite{Essler2001, Jeckelmann2003} and time-dependent DMRG (tDMRG)~\cite{Ohmura2019} for large systems of approximately 100 sites. 
Notably, a good agreement has been achieved, particularly in the case of ET-F$_2$TCNQ with parameters $(U, V) = (10, 3)$~\cite{Yamaguchi2021}.
We show the linear absorption spectrum Re$[\sigma(\omega)]$ of 1DEHM with $(U, V) = (10, 3)$ obtained by tDMRG in Fig.~\ref{Fig1}.
The inclusion of $V$ is essential for obtaining Re$[\sigma(\omega)]$ comparable with experiments in 1D Mott insulators, as an excitonic absorption peak induced by $V$ is clearly observed at $\omega=6$.

Moreover, it is known that excitonic levels with even and odd parities are almost degenerate, but the level with odd parity is slightly lower than that with even parity~\cite{Mizuno2000, Yamaguchi2021}. 
As a result, the transition dipole moment between them is anomalously large~\cite{Mizuno2000}.
We will demonstrate later that nonlinear optical processes using the excitonic levels influence the transient absorption spectra.

We investigate photoexcited states of 1DEHM with $\Omega=0$, 3, 6, and 8 pulses, whose shapes and energies are depicted in Fig.~\ref{Fig1}.
The photon energy of an $\Omega=0$ pulse is too small to be absorbed.
We note here that the central frequency of the $\Omega=0$ pulse used in this paper is not exactly zero, but about 0.3.
Since at $\Omega=3$ two photons are absorbed into an excitonic level at $\omega=6$, the Keldysh crossover is situated between $\Omega=0$ and 3.

\section{\label{sec:level3}Method: time-dependent DMRG}
Before presenting our results, we provide a brief overview of our numerical approach, which leads to time-dependent optical conductivity.
DMRG~\cite{White1992} is a kind of tensor network method representing wave functions, which are generally expressed as tensors, in terms of matrix products. 
The method is based on variational principles, involving the local singular value decomposition of matrices. 
In an optimization process, a low-rank approximation is achieved by discarding small singular values beyond the $\chi$-th value, where the singular values correspond to the eigenvalues of the reduced density matrix. 
The parameter introduced here, $\chi$, is referred to as the bond dimension, determining the accuracy of DMRG. 
In 1D systems that satisfy the area law of entanglement entropy, small $\chi$ is sufficient for obtaining an accurate ground state.

According to the time-dependent Schr\"{o}dinger equation, the evolved wave function at time $t$ from a certain initial state $|\psi(0)\rangle$ follows $|\psi(t)\rangle = U(t,0)|\psi(0)\rangle$ with
\begin{align}
    U(t,0)=\hat T \exp \left[ -i\int _0 ^t ds H(s) \right],
\end{align}
where $H(t)$ represents the time-dependent Hamiltonian, and $\hat T$ is the time-ordering operator.
Since $H(t)$ is generally non-commutative at different times, handling time-ordered products is complicated.
The complexity can be avoided by employing a small time step $\Delta t$ and assuming that the time variation of $H(t)$ within $\Delta t$ is negligible.
This assumption renders the approximation $U(t+\Delta t,t)=\exp[-i\Delta tH(t)]$ valid.
Consequently, it is only necessary to precisely compute the operator $\exp[-i\Delta tH(t)]$ to obtain accurate time-evolved states.

One commonly used method for obtaining $\exp[-i\Delta tH(t)]$ is based on the Suzuki-Trotter decomposition, which is a technique widely employed in 1D systems~\cite{White2004}.
However, when long-range interactions are present, as in the case of tDMRG introducing many long-range interactions to construct two-dimensional systems, the application of this method becomes considerably complicated, particularly due to the requirement for swap-gate operations.

To obtain $\exp[-i\Delta tH(t)]$, we take another approach with a polynomial as
\begin{align}
    U(t+\Delta t,t) 
            \simeq \sum_{l=0}^{M_\text{p}} (-i)^l (2l+1)\mathfrak{j}_l(\Delta t)P_l[\mathcal{H}(t)].
\end{align}
The Legendre polynomial $P_l(s)$ and spherical Bessel function of the first kind $\mathfrak{j}_l(s)$ are efficiently obtained using the recurrence relations
\begin{align}
P_0(x)=&1, \\
P_1(x)=&x, \\
    P_{l+1}(x) =& \frac{2l+1}{l+1}xP_l(x) - \frac{l}{l+1}P_{l-1}(x),
\end{align}
and
\begin{align}
\mathfrak{j}_0(x)=&x^{-1}\sin x, \\
\mathfrak{j}_1(x)=&x^{-1}(-\cos x + x^{-1}\sin x), \\
    \mathfrak{j}_{l+1}(x) =& (2l+1)x^{-1}\mathfrak{j}_l(x) - \mathfrak{j}_{l-1}(x),
\end{align}
respectively.
In practice, $M_\text{p} \sim 10$ with a time step of $\Delta t=0.02$ typically yields sufficient convergence. 
Additionally, to efficiently construct the bases spanned in time-dependent Hilbert space, our tDMRG is optimized for two target states $|\psi(t) \rangle$ and $|\psi(t+\Delta t) \rangle$. 
The multi-target approach combined with the Legendre polynomial expansion of a time-evolution operator is very efficient for obtaining time evolution of many-body wave functions not only in 1D but also in two-dimensional systems with high accuracy~\cite{Ohmura2019, Shinjo2021, Shinjo2021b, Shinjo2022, Shinjo2023, Tohyama2023}.
While a single-target approach is possible, but we should more carefully check convergences with respect to $\Delta t$ and $\chi$.

The optical conductivity can be calculated from the linear response of the current $j_\text{probe}(t)$ to a small probe electric field~\cite{Shao2016, Shinjo2018}, i.e.,
\begin{align}
    \sigma (\omega)=\frac{j_\text{probe}(\omega)}{i(\omega+i\gamma)LA_\text{probe}(\omega)},
\end{align}
Here, spatially homogeneous electric field $E(t) = -\partial A(t)$ is incorporated via the Peierls substitution in the hopping terms with the vector potential $A(t)$. 
Also, $A_\text{probe}(\omega)$ and $j_\text{probe}(\omega)$ are Fourier transform of 
$A_\text{probe}(t)=A_{0}^\text{pr}e^{-(t-t_0^\text{pr})^{2}/(\sqrt{2}t_\text{d}^\text{pr})^{2}}\cos [\Omega^\text{pr} (t-t_0^\text{pr})]$
and
$j_\text{probe}(t)=\langle \partial{H(t)}/\partial{A_\text{probe}(t)} \rangle$, respectively.
$\gamma$ is introduced as a broadening factor.
In this paper, we take $A_{0}^\text{pr}=0.001$, $t_\text{d}^\text{pr}=0.02$, and $\Omega^\text{pr}=10$.

The linear absorption spectrum is independent of $t_0^\text{pr}$, but when a pump pulse with a vector potential $A(t)=A_{0}e^{-(t-t_0)^{2}/(\sqrt{2}t_\text{d})^{2}}\cos [\Omega (t-t_0)]$ is introduced, $\tau=t_0^\text{pr}-t_0$ represents the delay time of the probe pulse relative to the pump pulse, and the transient absorption spectrum is expressed as $\sigma(\omega,\tau)$.
In this paper, we use pump pulses with $t_d=2$ and $t_0=10$, and the electric-field amplitude $E_0$ of a pump pulse is obtained with the relation $E(t)=-\partial A(t)$.

\begin{figure*}[t]
\includegraphics[width=0.95\textwidth]{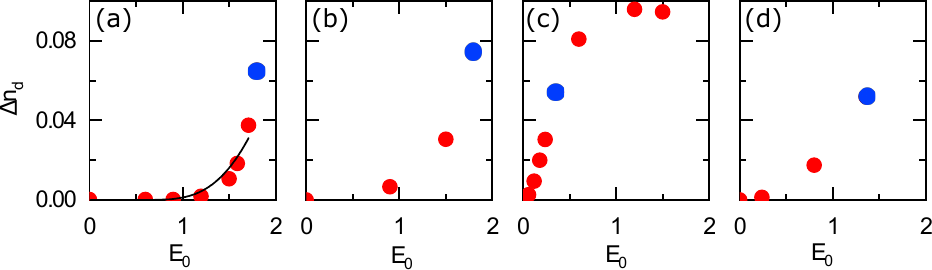}
\caption{$\Delta n_\text{d}$ of the $L = 32$ half-filled 1DEHM for $(U, V ) = (10, 3)$ excited by electric pulses with (a) $\Omega=0$, (b) $\Omega=3$, (c) $\Omega=6$, and (d) $\Omega=8$. The black line shows a fitted curve proportional to $E_0 \exp (-\pi E_\text{th}/E_0)$. $E_0$ indicated by the blue points is used in Fig.~\ref{Fig3}. Taken from Ref.~\onlinecite{Shinjo2022} with modifications.}
\label{Fig2}
\end{figure*}

\section{Keldysh crossover}

In this section, we demonstrate that the changes in doublon density, Drude response, and entanglement entropy due to a pump pulse are qualitatively different on the two sides across the Keldysh crossover.

\subsection{Photo-doped carriers}

We show in Fig.~\ref{Fig2} how the Keldysh crossover appears as a change in photo-doped carrier density.
Pump pulses induce the changes of doublon density in 1DEHM, $\Delta n_\text{d}=\frac{1}{L}\bigl[ \overline{\langle I \rangle_{t}} - \langle I\rangle_0 \bigr]$, shown in Fig.~\ref{Fig2}, where $I=\sum_{j}n_{j,\uparrow}n_{j,\downarrow}$, $ \overline{\langle \mathcal{O} \rangle_{t}}$ is the average of an expectation value of an operator $\mathcal{O}$ from $t= \tau+t_0-1$ to $\tau+t_0$, just before a probe pulse is applied, and $\langle \mathcal{O} \rangle_0$ is an expectation value of $\mathcal{O}$ for a ground state.
When we produce doublons by an $\Omega=0$ monocycle pulse, $\Delta n_\text{d}$ is not proportional to $E_0$ even for small $E_0$, but follows a threshold behavior~\cite{Oka2012} $\Delta n_\text{d} \propto E_{0}\exp \left(-\pi E_\text{th}/E_{0} \right)$ with a threshold field $E_\mathrm{th}$, as indicated by the black line in Fig.~\ref{Fig2}(a).
This threshold behavior is a distinctive feature of quantum tunneling and has recently been observed~\cite{Yamakawa2017, Li2022, Takamura2023}.

In contrast, when doublons are induced by $\Omega=3$, 6, and 8 multicycle pulses, the change in doublon density $\Delta n_\text{d}$ is found to be proportional to $E_0$ for small values of $E_0$.
An $\Omega=8$ multicycle pulse excites electrons in a continuum leading to $\Delta n_\text{d}\propto E_{0}^{2}$ as discussed in Ref.~\onlinecite{Oka2012}.
Since $\Omega=6$ resonates with the energy of an excitonic level, doublons are efficiently produced even at a small electric field through a one-photon absorption process.
On the other hand, $\Omega=3$ resonates with the half energy of the excitonic level, and doublons are produced by a two-photon absorption process.
In any case, no threshold behavior is seen when doublons are excited by photon absorption.

\subsection{\label{sec:level4}Transient absorption spectra}
\begin{figure*}[t]
\includegraphics[width=1\textwidth]{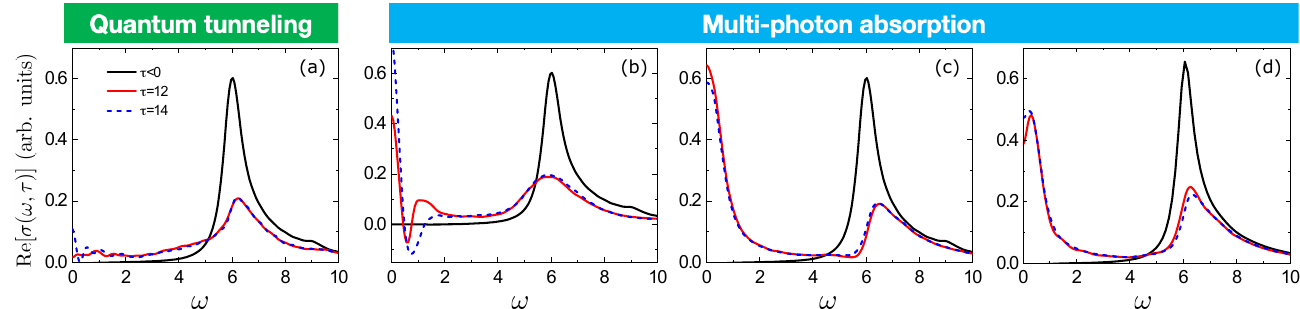}
\caption{Re$\sigma(\omega,\tau)$ of half-filled 1DEHM excited by (a) an $\Omega=0$ monocycle pulse, (b) $\Omega=3$ (c) $\Omega=6$, and $\Omega=8$ multicycle pulses. Black, red, and blue-dashed lines are for $\tau<0$, $\tau=12$, and 14, respectively. $L=32$ is used in (a)--(c), but $L=48$ in (d). $\gamma=0.4$ is used. $E_0$ is taken from the blue points in Fig.~\ref{Fig2}.}
\label{Fig3}
\end{figure*}
\begin{figure}[htb]
\includegraphics[width=0.4\textwidth]{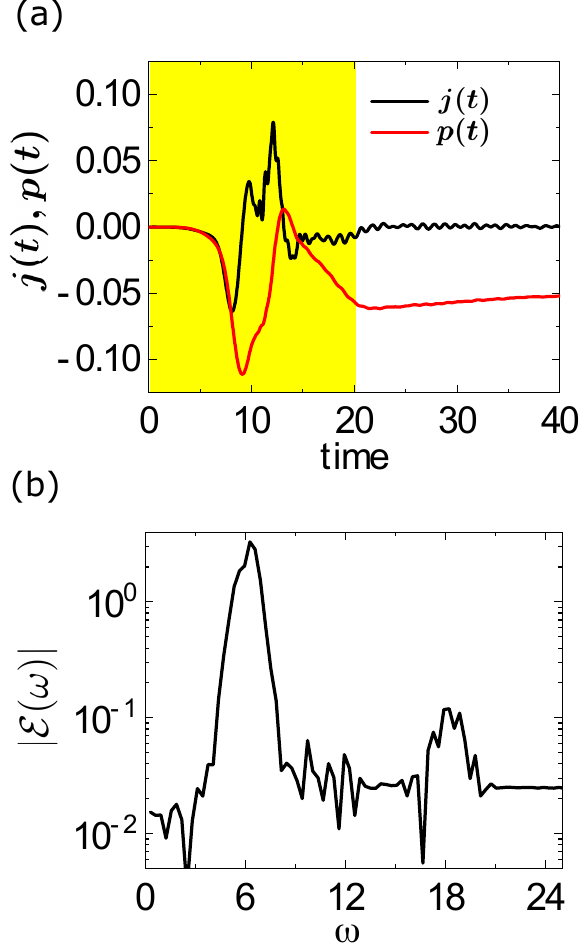}
\caption{(a) $j(t)$ and $p(t)$, denoted by black and red lines, respectively, for the half-filled 1DEHM with $(U,V)=(10,3)$ on $L=32$ after the initial ground state at $t=0$ is excited by a monocycle $\Omega=0$ pulse with $t_\text{d}=2$ and $E_{0}=2$ applied at $t_{0}=10$. 
    The yellow shade is a guide to the eye to indicate the time during which the pulse is applied.
    (b) Harmonic generation $|\mathcal{E}(\omega)|$ for the photoexcited state prepared in (a) evaluated from $j(t)$ in $20<t<40$ that is induced by an $\Omega=6$ pulse with $t_\text{d}=2$ and $E_{0}=1$ applied subsequently at $t_{0}=30$. Taken from Ref.~\onlinecite{Shinjo2023}.}
\label{Fig_shg}
\end{figure}

In Fig.~\ref{Fig3}, we show how the Keldysh crossover appears as a change in transient absorption spectra.
Figure~\ref{Fig3}(a) is Re$[\sigma(\omega,\tau)]$ of 1DEHM excited by quantum tunneling with a strong $\Omega=0$ monocycle pulse.
The central frequency of the $\Omega=0$ monocycle pulse is 0.3, which is too small to excite the Mott gap.
$E_0=1.8$ used to obtain Fig.~\ref{Fig3}(a) leads to $\Delta n_\text{d}=0.07$, which corresponds to the blue point in Fig.~\ref{Fig2}(a).
Since spin-charge separation is realized in 1DEHM, the spectral weights above the Mott gap transfer to lower energies at $\omega=0$, i.e., the Drude weight $\sigma_\text{D}$.
We find that $\sigma_\text{D}$ in Fig.~\ref{Fig3}(a) is not proportional to $\Delta n_\text{d}$ but is strongly suppressed~\cite{Shinjo2022,comment1}.
Here, we note that the Drude weight appears at $\omega > 0$ due to a finite-size effect and its peak approaches $\omega=0$ as $L$ increases.
For $L=32$, we can mask this finite-size effect by taking $\gamma = 0.4$.

We note that the absence of a Mott insulator-to-metal transition is consistent with dynamical mean-field theory~\cite{Eckstein2010}, but inconsistent with experimental observations~\cite{Yamakawa2017, Takamura2023}.
One of the reasons for the discrepancy with the experiments may be that our system is an isolated electron system, which does not incorporate electron-phonon couplings.
Since localized properties such as many-body localization are very weak against dissipation, the glassy nature of the excited states induced by the $\Omega=0$ monocycle pulse can be easily destroyed by introducing phonon degrees of freedom.

We also note that an absorption induced at $\omega \simeq 4$ is found in Fig.~\ref{Fig3}(a).
We consider that this structure is due to the Stark shift of the excitonic level, and may correspond to a structure at 0.6eV observed in ET-F$_2$TCNQ\cite{Takamura2023}.

The strong suppression of $\sigma_\text{D}$ suggests that strong fields localize nonequilibrium states.
The electric field in the low-frequency limit can be regarded as a tilted potential, which causes resonant quantum tunneling of electrons.
As a result, doublons and holons are produced accompanying polarization. 
We can derive an effective model~\cite{Shinjo2022} $\mathcal{H}_\text{eff}$ in a large $U$ Hubbard model with a tilted potential, which effectively conserves the sum of polarization $P=\sum_{j}jn_{j}$ and doublon $I=\sum_{j}n_{j,\uparrow}n_{j,\downarrow}$ as $[\mathcal{H}_\text{eff},P+I]=0$.
This conservation imposes a constraint on the kinetic motion of electrons~\cite{Scherg2021}, maintaining polarization when the doublon density does not change.

As shown in Fig.~\ref{Fig_shg}(a), the polarization induced by applying an $\Omega=0$ monocycle pulse in 1DEHM persists after pulse irradiation.
This means that a photo-excited state undergoes a change that breaks inversion symmetry with finite polarization $p(t)=\int_{0}^{t}ds j(s)$~\cite{Shinjo2023}.
As a result of inversion-symmetry breaking, we find second harmonic generation at $\omega \simeq 12$ induced by a pulse with $\Omega=6$ in Fig.~\ref{Fig_shg}(b), where harmonic generation defined as $\mathcal{E}(\omega)=\omega \tilde{j}(\omega)$ evaluated from $\tilde{j}(\omega)=\frac{1}{2\pi}\int dt e^{i\omega t}j(t)$ is plotted.
In the non-resonant case, the restriction of electron motion can also be imposed under a strong electric field, but only while the electric field is present.
Because of a transient phenomenon just after the change of electronic states by resonance quantum tunneling and the end of direct doublon generation by the pulse, the induced polarization does not immediately vanish after the electric field disappears, but gradually decreases.
Further investigations of high-harmonic generation~\cite{Silva2018, Murakami2018, Murakami2021} remain as future work, which can give additional information on electronic states.

In contrast to quantum tunneling, photon absorption induces a metallic state with a large $\sigma_\text{D} \propto \Delta n_\text{d}$ in 1DEHM. 
We show Re$\sigma(\omega,\tau)$ of 1DEHM induced by photon absorption in Figs.~\ref{Fig3}(b)--\ref{Fig3}(d).
The excitonic level at $\omega=6$ is excited via two-photon and one-photon absorption processes with $\Omega=3$ [see Fig.~\ref{Fig3}(b)] and $\Omega=6$ [see Fig.~\ref{Fig3}(c)] photon pulses, respectively.
The continuum followed by the excitonic level is excited by an $\Omega=8$ photon pulse in Fig.~\ref{Fig3}(d).
The electric fields $E_0=1.8$, 0.36, and 1.4 used to obtain Figs.~\ref{Fig3}(b), \ref{Fig3}(c), and \ref{Fig3}(d) lead to $\Delta n_\text{d}=0.07$, 0.05, and 0.05 which correspond to the blue points in Figs.~\ref{Fig2}(b), \ref{Fig2}(c), and \ref{Fig2}(d), respectively.
In all cases, we find a clear enhancement of the Drude weight proportional to photo-doped carrier density~\cite{Shinjo2022, comment1}, i.e., $\sigma_\text{d} \propto \Delta n_\text{d}$.
The Mott insulator-to-metal transition due to photon absorption is consistent with the observation in 1D Mott insulators.~\cite{Taguchi2000, Iwai2003, Okamoto2007}

In addition to the Drude response, we find spectral weights at $\omega=0.3$, which are negative and positive in Figs.~\ref{Fig3}(b) and \ref{Fig3}(c), respectively.
These structures are induced by stimulated emission and absorption between nearly-degenerated excitonic levels with a gap of 0.3.
We note that since the continuum has a discrete structure in a finite system, Re$[\sigma(\omega,\tau)]$ is sensitive to finite-size effects when 1DEHM is excited with an $\Omega=8$ pulse.
This is the reason why we use a $L=48$ cluster in Fig.~\ref{Fig3}(d).

\subsection{\label{sec:level5}Entanglement entropy}
\begin{figure}[t]
\includegraphics[width=0.45\textwidth]{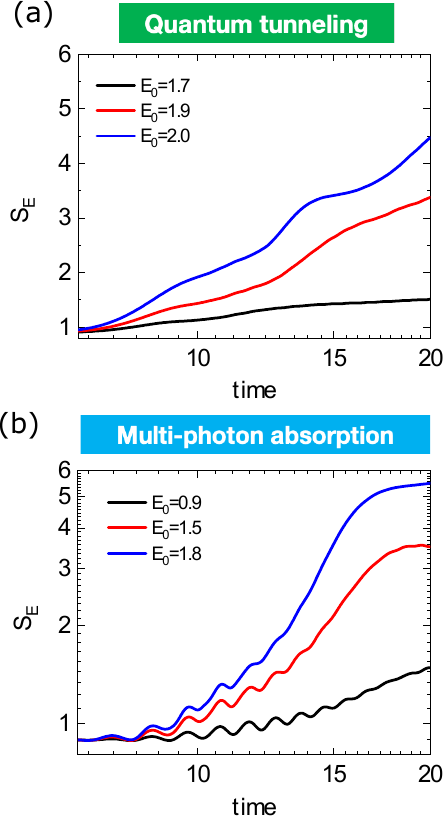}
\caption{The time evolution of $S_\text{E}$ in the half-filled $L=32$ 1DEHM excited by (a) an $\Omega=0$ monocycle pulse and (b) an $\Omega=3$ pulse, which induce quantum tunneling and photon absorption, respectively. Note that (a) semi-log plot and (b) log-log plot are used.}
\label{Fig5}
\end{figure}

The time evolution of an entanglement entropy $S_\text{E}=-\sum_{i}p_{i}\ln p_{i}$ shows a contrasting behavior between quantum tunneling and photon absorption.
Here, $p_{i}$ is obtained by the Schmidt decomposition of a wavefunction $|\psi \rangle$ as
\begin{align}
|\psi \rangle = \sum_{i} p_{i} |\psi_A^{i}\rangle |\psi_B^{i}\rangle,
\end{align}
where a system is composed of two subsystems $A$ and $B$.
We obtain $S_\text{E}$ by making $A$ half of the whole system.
In Fig.~\ref{Fig5}(a), we show the time evolution of $S_\text{E}$ in an $L=32$ half-filled 1DEHM excited by an $\Omega=0$ monocycle pulse, which induces quantum tunneling.
We find that $S_\text{E}$ shows a slow logarithmic growth.
For comparison, we show in Fig.~\ref{Fig5}(b) the time evolution of $S_\text{E}$ when 1DEHM is excited with an $\Omega=3$ pulse, whose photons are absorbable.
We see that $S_\text{E}$ shows a rapid linear growth and then saturates at the end of pulse irradiation, i.e., $t=20$.
Since the initial state is not a product state with $S_\text{E}= 0$ and the increase in $S_\text{E}$ depends on the amplitudes of electric fields $E_{0}$, underlying physics is nontrivial and complicated.
However, we can see a tendency for $S_\text{E}$ to grow more slowly for quantum tunneling than for photon absorption.
The slow growth of $S_\text{E}$~\cite{Znidaric2008, Bardarson2012, Serbyn2013} is considered to be one of the manifestations of the localized nature of excited states induced by a strong terahertz pulse.

\section{\label{sec:level6}Summary}
Using tDMRG, we have investigated the Keldysh crossover in 1DEHM with $(U,V)=(10,3)$ excited by pump pulses with frequency $0 < \Omega \leq 8$.
Investigating time-dependent optical conductivity Re$\sigma (\omega,\tau)$ of the half-filled 1DEHM, we have found that the enhancement of the Drude weight with $\sigma_\text{D} \propto \Delta n_\text{d}$ is induced by photon absorption with $\Omega=3$, 6, and 8 multicycle pulses.
In contrast, we have found no enhancement of the Drude weight even for photo doping by quantum tunneling with an $\Omega=0$ monocycle pulse.
As a result, we conclude that photon absorption leads to a Mott insulator-to-metal transition, but quantum tunneling leads to a Mott insulator-to-glass transition, which is a manifestation of the Keldysh crossover seen in optical properties.
Due to the localized nature of nonequilibrium states with an $\Omega=0$ monocycle pulse, $S_\text{E}$ shows a logarithmic growth.

In ET-F$_2$TCNQ, pump-probe spectra have been observed with excitations at $\Omega=0.01$eV~\cite{Takamura2023}, 0.69eV~\cite{Miyamoto2019}, and 1.55eV~\cite{Uemura2008}. 
We expect to observe the Keldysh crossover in 1D Mott insulators by further experiments with various photon energies.

\begin{acknowledgments}
We acknowledge discussions with S. Sota, and S. Yunoki. This work was supported by CREST (Grant No. JPMJCR1661), the Japan Science and Technology Agency, by the Japan Society for the Promotion of Science and KAKENHI (Grants No. 19H05825) from Ministry of Education, Culture, Sports, Science, and Technology (MEXT), Japan. The numerical calculation was carried out using computational resources of HOKUSAI at RIKEN Advanced Institute for Computational Science, the supercomputer system at the information initiative center, Hokkaido University, the facilities of the Supercomputer Center at Institute for Solid State Physics, the University of Tokyo, and supercomputer Fugaku provided by the RIKEN Center for Computational Science through the HPCI System Research Project (Project ID: hp220048, hp230066).
\end{acknowledgments}

\section*{Data Availability Statement}
Data supporting the findings of this study are available from the authors upon reasonable request.


\end{document}